\documentclass[12pt]{article}
\usepackage{amssymb,amsmath,epsfig}

\begin{document}

\title{\bf Complex Wave Numbers in the Vicinity of the Schwarzschild Event Horizon}

\author{M. Sharif \thanks{msharif@math.pu.edu.pk} and Umber Sheikh\\
%EndAName
{\small Department of Mathematics,}\\
{\small University of the Punjab, Lahore 54590, Pakistan}}

\date{}
\maketitle
\begin{abstract}
This paper is devoted to investigate the cold plasma wave
properties outside the event horizon of the Schwarzschild planar
analogue. The dispersion relations are obtained from the
corresponding Fourier analyzed equations for non-rotating and
rotating, non-magnetized and magnetized backgrounds. These
dispersion relations provide complex wave numbers. The wave
numbers are shown in graphs to discuss the nature and behavior of
waves and the properties of plasma lying in the vicinity of the
Schwarzschild event horizon.
\end{abstract}

{\bf Keywords }: Schwarzschild planar analogue, cold plasma.

\section{Introduction}

General Relativity is the geometric theory of gravity which deals
with four- dimensional spacetime. According to this theory, gravity
severely modifies space and time near the black hole. The
Schwarzschild black hole, a minimum configuration of the Kerr black
hole, exhibits a strong gravitational field \cite{P}. This field
pulls the plasma surrounding the black hole event horizon towards
and then into the black hole. The plasma is drawn equally from all
directions and thus form an omni-directional accretion disk. The
existence of this disk is one of the observational facts used to
identify a black hole. The motion of plasma inside the accretion
disk is governed by the theory of general relativistic
magnetohydrodynamics (GRMHD). Maxwell's equations, Ohm's law, mass,
momentum and energy conservation equations constitute this theory
that are required to investigate various aspects of the interaction
of relativistic gravity with plasma's magnetic field.

The theory of general relativity is a sterile subject until it
touches the real physical world. Only the hard reality of
experiments and of astronomical observations can bring the theory to
life. The 3+1 ADM formalism, developed to study the quantization of
gravitational field by Arnowitt et al. \cite{ADM}, has mostly been
used in numerical relativity \cite{SJ}-\cite{ESW}. The 3+1 spacetime
split in the formulation of general relativity is particularly
appropriate for applications to the black hole theory as described
by Thorne et al. \cite{TM1}-\cite{TPM}. Wave properties in the
Friedmann universe were investigated \cite{HT}-\cite{De} by the same
formalism. Sakai and Kawata \cite{SK} developed a linearized
treatment of relativistic plasma waves in analogy with the special
relativistic formulation. Khanna \cite{Kh} derived Ohm's law for two
component plasma theory of the Kerr black hole. Zhang
\cite{Z1}-\cite{Z2} formulated stationary symmetric GRMHD theory
with its applications in Kerr geometry. Buzzi et al.
\cite{BH1}-\cite{BH2} treated the wave propagation in radial
direction close to the Schwarzschild horizon in general relativistic
two component plasma.

In a recent paper, Sharif and Umber \cite{U} have investigated real
wave numbers and properties of the medium existing in the vicinity
of a Schwarz-schild black hole using the same split. The cold plasma
is considered in non-rotating or rotating, non-magnetized or
magnetized backgrounds. The same authors have also worked out the
wave properties in isothermal plasma for the Schwarzschild spacetime
planar analogue using ADM split \cite{S1}-\cite{S2}. It is verified
that no information can be extracted either from the event horizon
or its exterior.

In this paper, we focus our attention to investigate the wave
numbers and properties of the medium for the Schwarzschild planar
analogue spacetime. The paper is organized as follows. Section 2
describes the general line element and its specified form (the
planar analogue) for the Schwarzschild black hole in 3+1 formalism.
It also contains background assumptions. In sections 3, 4 and 5, the
dispersion relations for non-rotating (non-magnetized and
magnetized), rotating non-magnetized and rotating magnetized
backgrounds are investigated respectively. Finally, section 6
provides the outlook of the results.

\section{3+1 Spacetime Modeling}

The general line element in $3+1$ formalism can be expressed as
\cite{Z2}
\begin{equation*}\label{a}
ds^2=-\alpha^2dt^2+\gamma_{ij}(dx^i+\beta^idt)(dx^j+\beta^jdt),
\end{equation*}
where $\alpha$ is the lapse function, $\beta^i$ are the components
of the shift vector and $\gamma_{ij}$ are the components of the
spatial metric. All these quantities are functions of coordinates
$t$ and $x^i$. A natural observer, associated with this spacetime
called fiducial observer (FIDO), has four-velocity perpendicular to
the hypersurfaces of constant time $t$. The planar analogue of the
Schwarzschild line element \cite{Z2} is
\begin{eqnarray}{\setcounter{equation}{1}}\label{p}
ds^2=-\alpha^2(z)dt^2+dx^2+dy^2+dz^2,
\end{eqnarray}
where $z$, $x$ and $y$ are analogues of radial $r$, $\phi$ and
$\theta$ directions respectively. The lapse function vanishes at
the event horizon placed at $z=0$, and increases monotonically to
unity as $z$ increases from $0$ to $\infty$.

\subsection{Background Assumptions}

We consider two backgrounds with the specifications given below:
\begin{enumerate}
\item \textbf{Non-Rotating Background}

In this background, the fluid flow as well as the external magnetic
field lines are along $z$-direction (analogous to the radial
direction) moving towards the black hole event horizon. The fluid
velocity and the magnetic field lines are taken to be
$\textbf{V}=u(z)\textbf{e}_\textbf{z}$ and
$\textbf{B}=B\textbf{e}_\textbf{z}$ respectively.
\item \textbf{Rotating Background}

This background demands the fluid to move in more than one
dimension. Thus we assume that the fluid motion and the external
magnetic field according to FIDO  as
$\textbf{V}=V(z)\textbf{e}_\textbf{x}+u(z)\textbf{e}_\textbf{z}$ and
$\textbf{B}=\lambda(z)
B\textbf{e}_\textbf{x}+B\textbf{e}_\textbf{z}$ respectively, where
$B$ is a constant.
\end{enumerate}

In MHD theory, plasma is treated to be a perfect fluid. We assume
cold plasma surrounding the Schwarzschild black hole with equation
of state
\begin{equation}
\mu=\frac{\rho}{\rho_0}=constant
\end{equation}
which has vanishing thermal pressure and thermal energy. Here $\rho$
is the mass density of the fluid. The fluid flow is assumed to be
perturbed by the gravitational field of the black hole. The
linearized perturbed quantities take the form
\begin{eqnarray}\label{b}
\rho=\rho^0+\rho \tilde{\rho},\quad
\textbf{V}=\textbf{V}^0+\textbf{v},\quad
\textbf{B}=\textbf{B}^0+B\textbf{b}.
\end{eqnarray}
The variables with superscript zero are unperturbed quantities. The
dimensionless notations for the perturbed quantities are
\begin{eqnarray}\label{b1}
\tilde{\rho}=\tilde{\rho}(t,z),\quad
\textbf{v}=v_z(t,z)\textbf{e}_\textbf{z},\quad
\textbf{b}=b_z(t,z)\textbf{e}_\textbf{z}
\end{eqnarray}
for non-rotating background whereas for rotating background
\begin{eqnarray}\label{b4}
\textbf{v}=v_x(t,z)\textbf{e}_\textbf{x}+v_z(t,z)\textbf{e}_\textbf{z},\quad
\textbf{b}=b_x(t,z)\textbf{e}_\textbf{x}+b_z(t,z)\textbf{e}_\textbf{z}.
\end{eqnarray}

We assume that the perturbed quantities have harmonic space and time
dependence and hence these quantities can be expressed as follows
\begin{eqnarray}\label{b6}
&&\tilde{\rho}(t,z)=c_1e^{-\iota (\omega t-kz)},\quad
v_z(t,z)=c_2e^{-\iota (\omega t-kz)},\quad
v_x(t,z)=c_3e^{-\iota (\omega t-kz)},\nonumber\\
&&b_x(t,z)=c_4e^{-\iota (\omega t-kz)},\quad b_z(t,z)=c_5e^{-\iota
(\omega t-kz)}
\end{eqnarray}
where $c_1,~c_2,~c_3,~c_4$ and $c_5$ are arbitrary constants, $k$
is the wave number and $\omega$ is the angular frequency of waves.

\section{Non-Rotating Background}

For the fluid flow in only one dimension, i.e., $z$-direction, we
shall use the same Fourier analysed equations given in \cite{U} as
follows
\begin{eqnarray}{\setcounter{equation}{1}}
\label{p1}
&&-\frac{\iota \omega}{\alpha} c_5=0,\\
\label{p2}
&&\iota kc_5=0,\\
\label{p3} &&c_1\left(\frac{-\iota \omega}{\alpha}+\iota k
u\right)+c_2\left\{(1+\gamma^2 u^2)\iota
k-(1-2\gamma^2u^2)\right.\nonumber\\
&&\left.\times(1+\gamma^2u^2)\frac{u'}{u}-\frac{i\omega}{\alpha}\gamma^2
u\right\}=0,\\
\label{p4} &&c_1\gamma^2 \{a_z+uu'(1+\gamma^2 u^2)\}
+c_2[\gamma^2(1+\gamma^2u^2)\left(\frac{-\iota \omega}{\alpha}
+\iota u k\right)+\nonumber\\
&&+\gamma^2\{u'(1+\gamma^2 u^2)(1+4\gamma^2 u^2)+2u\gamma^2
a_z\}]=0.
\end{eqnarray}
Equations (\ref{p1}) and (\ref{p2}) imply that $c_5=0$ which means
that no perturbations occur in magnetic field of the fluid. Thus
we are left with Eqs.(\ref{p3}) and (\ref{p4}). It is mentioned
here that for non-magnetized plasma, we also obtain the same two
equations.

\subsection{Numerical Solutions}

We consider cold plasma of constant density. The time lapse is
assumed to be $\tanh(z)$ which vanishes at the horizon $z=0$ and
increases to $1$ as $z\rightarrow\infty.$ Using mass conservation
law in three dimensions, i.e., $\rho \gamma u=constant$ we obtain
the value of $u=\frac{1}{\sqrt{1+\tanh^2(z)}}$. The determinant of
the coefficients of the constants $c_1$ and $c_2$ in Eqs.(\ref{p3})
and (\ref{p4}) is a complex number. Solving this determinant by
using the software \textit{Mathematica}, we obtain two complex
values of the wave number $k$. Real and imaginary parts of this wave
number show the propagation factor and the attenuation factor
respectively for plane wave in dispersive medium. The propagation
factor gives the phase velocity ($v_p=\frac{\omega}{Re(k)}$) and
group velocity ($v_g=\frac{d\omega}{d(Re(k))}$) of the waves. The
sinusoidal expressions then take the form $e^{-\iota(\omega t-k_1
z-\iota k_2z)}=e^{-\iota(\omega t-k_1 z)-k_2z}$, where $k_1=Re(k)$
and $k_2=Im(k)$. The wave numbers obtained are shown in Figures 1
and 2.

\begin{figure}
\center \epsfig{file=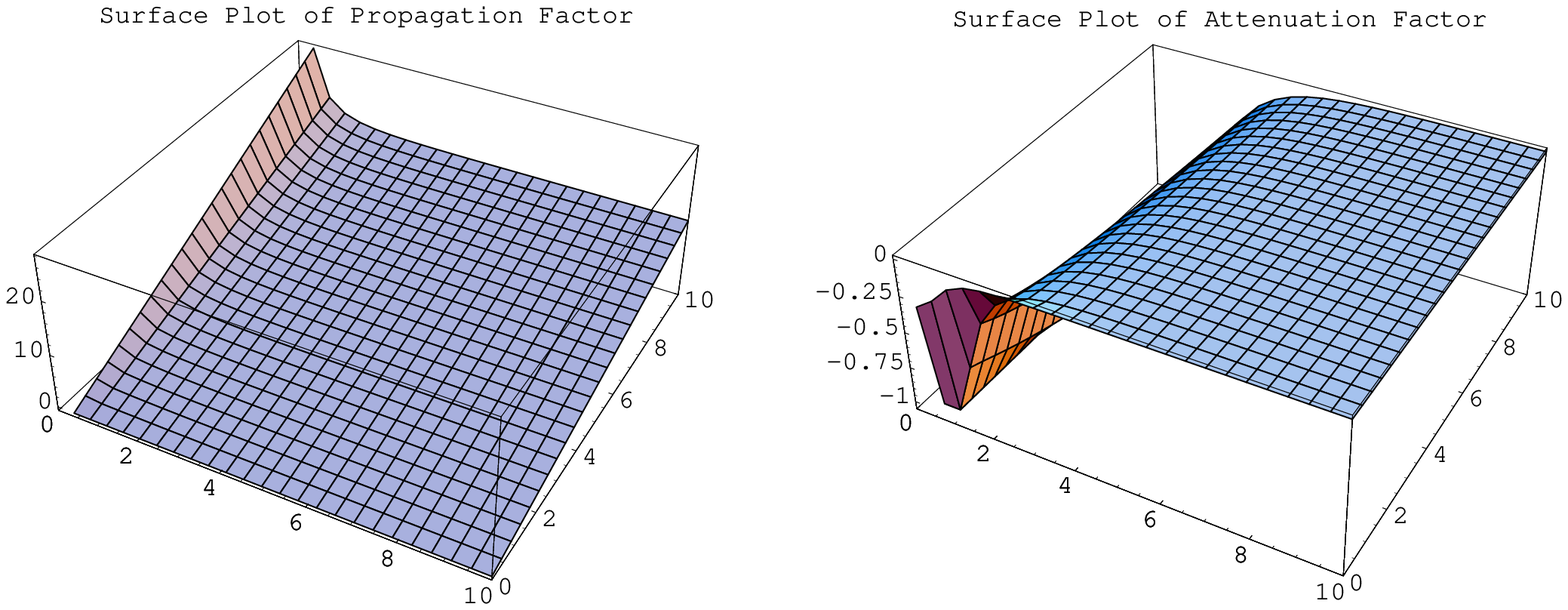,width=1.0\linewidth} \center
\epsfig{file=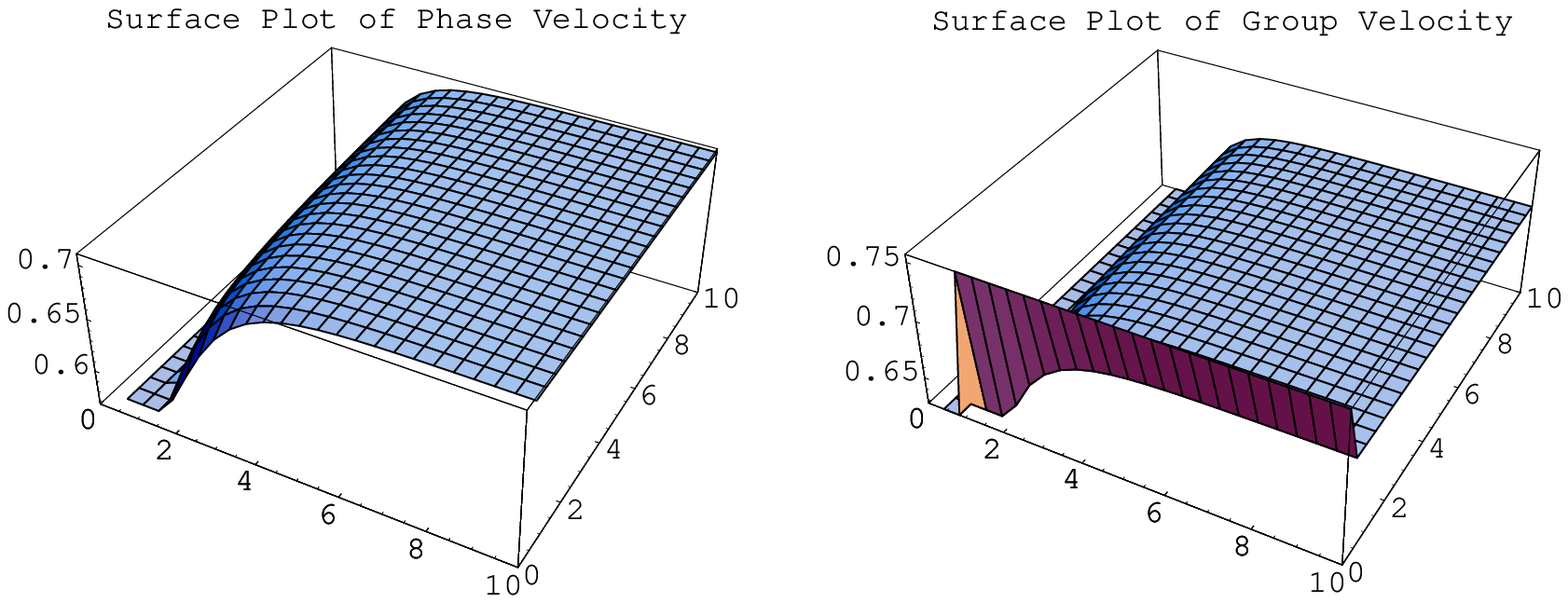,width=1.0\linewidth} \caption{Wave propagation
increases near the event horizon. Dispersion is found to be
anomalous.}
\end{figure}
In Figure 1, the waves vanish in the region $0\leq
z<1.1\times10^{-8}$ due to infinite wave number. The propagation
factor of the waves is positive for the region $0\leq z<1.1 \times
10^{-8},~0<\omega\leq10$. It increases with the increase in angular
frequency and decreases with an increase in $z$. Thus the waves
propagate rapidly far from the event horizon. The waves with higher
angular frequencies move faster than the waves with lower angular
frequencies. The attenuation factor increases with an increase in
$z$. The waves damp as they move away from the event horizon. The
phase velocity is less than the group velocity in the region
$0<z\leq10$ which implies anomalous dispersion of the waves in this
region \cite{Ac}. There is an exceptional case of the waves with
very small angular frequency that do not show this behavior.

\begin{figure}
\center \epsfig{file=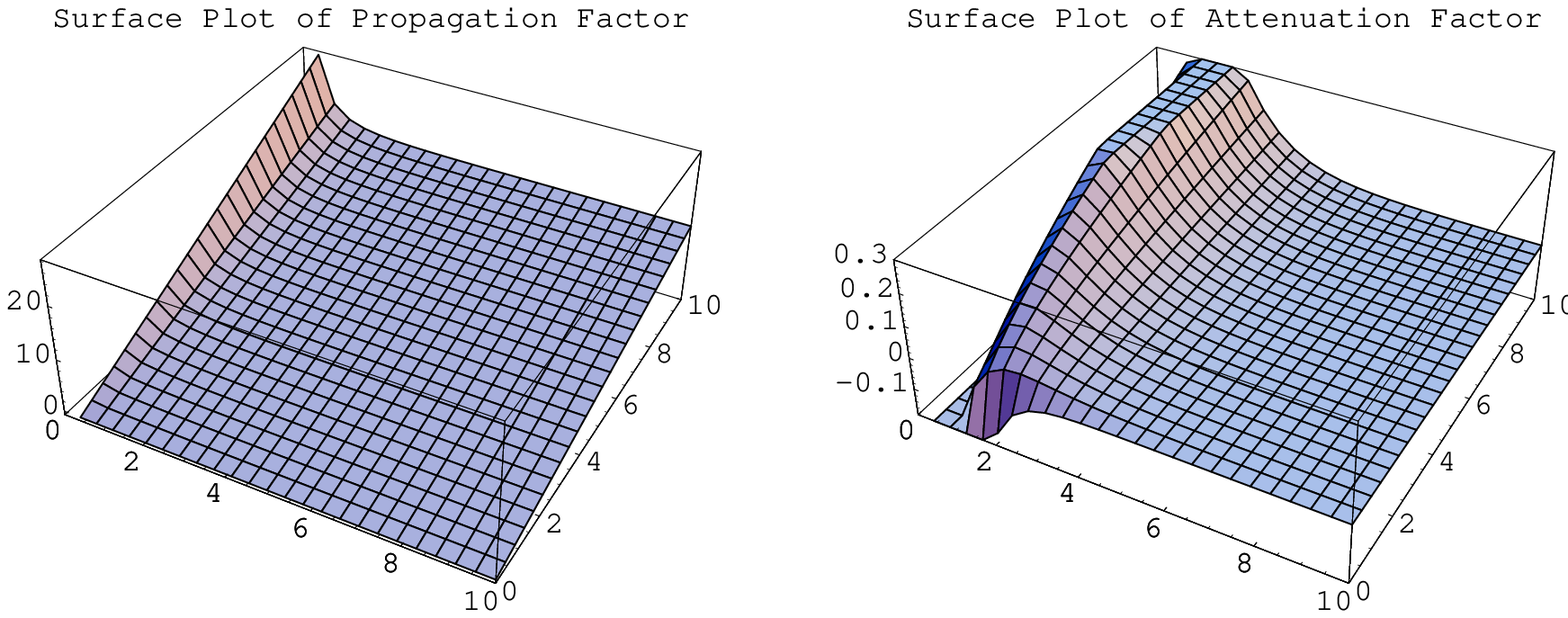,width=1.0\linewidth} \center
\epsfig{file=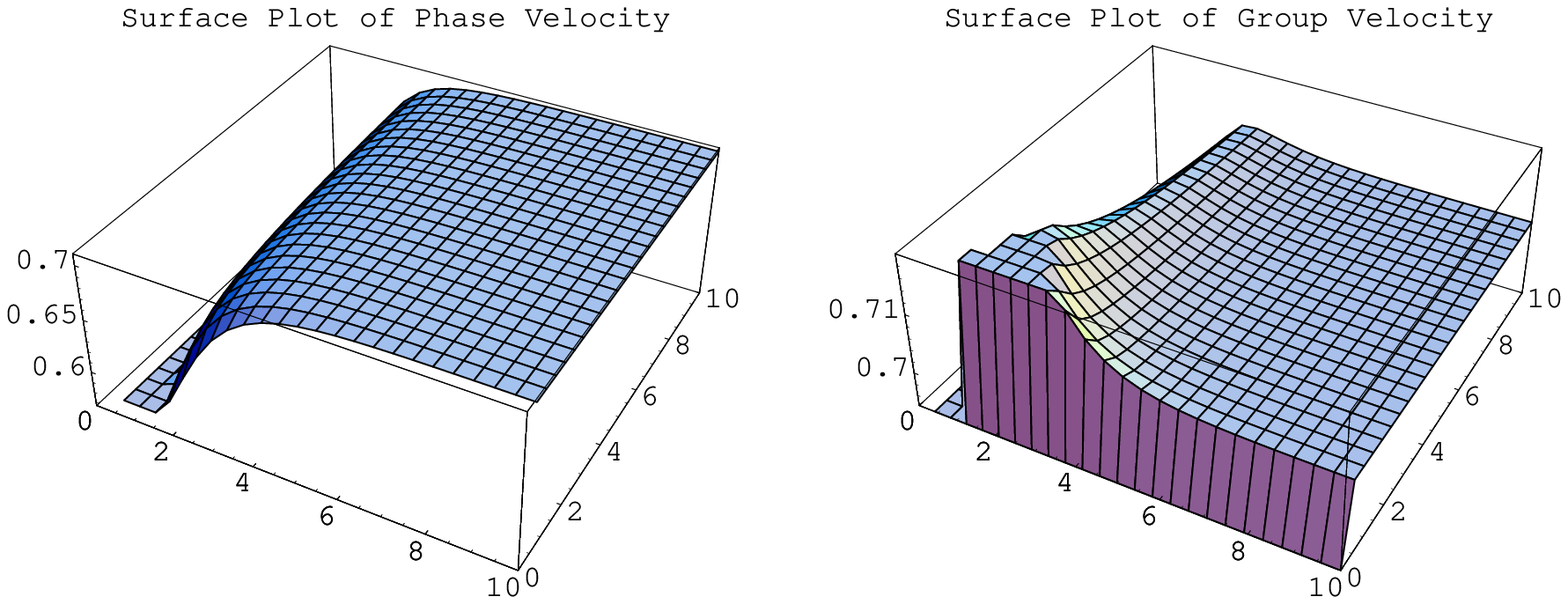,width=1.0\linewidth} \caption{The waves
propagate rapidly near the event horizon. The whole region admits
normal dispersion.}
\end{figure}
Figure 2 shows that the wave number is infinite in the region $0\leq
z<1.1\times10^{-8}$. The propagation factor takes negative values in
$1.1\times10^{-8}\leq z\leq10,~0\leq \omega<1\times10^{-19}$ which
shows the same behavior as given in Figure 1 except for this small
region. The attenuation factor is negative in the region
$1.1\times10^{-8}\leq z\leq 0.68$ and takes larger values at $z=1.4$
where the waves damp. The waves grow as the angular frequency
decreases. The group velocity is less than the phase velocity for
the region $0<\omega\leq 10$ which shows normal dispersion of waves.

\section{Rotating Non-Magnetized Background}

For the rotating background, the Fourier analysed Eqs.(4.13)-(4.15)
of \cite{U} are
\begin{eqnarray}{\setcounter{equation}{1}}\label{f4}
&&c_1\iota
\left(-\frac{\omega}{\alpha}+k u\right)+c_2\left[-\frac{\iota
\omega}{\alpha}\gamma^2u+(1+\gamma^2 u^2)\iota
k\right.\nonumber\\&&\left.-(1-2\gamma^2u^2)(1+\gamma^2u^2)\frac{u'}{u}+
2\gamma^4u^2VV'\right]\nonumber\\
&&+c_3\gamma^2\left[-\frac{\iota \omega}{\alpha}V+\iota
kuV+u\{(1+2\gamma^2V^2)V'+2\gamma^2uVu'\}\right]=0,\\
\label{f5}
&&c_1\gamma^2u\{(1+\gamma^2V^2)V'+\gamma^2uVu'\}+c_2\gamma^2[\iota\left(-\frac{
\omega}{\alpha}+ku \right)\gamma^2uV\nonumber\\
&&+\{(1+2\gamma^2u^2)(1+2\gamma^2V^2)-\gamma^2V^2\}V'
+2\gamma^2(1+2\gamma^2u^2)uVu']\nonumber\\
&&+c_3[\left(-\frac{\iota \omega}{\alpha}+\iota ku\right)
\gamma^2(1+\gamma^2V^2)+\gamma^4u\{(1+4\gamma^2V^2)uu'\nonumber\\
&&+4VV'(1+\gamma^2V^2)\}]=0,\\
\label{f6}
&&c_1\gamma^2\{a_z+(1+\gamma^2u^2)uu'+\gamma^2u^2VV'\}\nonumber\\
&&+c_2[\gamma^2(1+\gamma^2u^2)\iota\left(-\frac{
\omega}{\alpha}+uk\right)+\gamma^2\{u'(1+\gamma^2u^2)(1+4\gamma^2u^2)\nonumber\\
&&+2u\gamma^2(a_z+(1+2\gamma^2u^2)VV')\}]
+c_3\gamma^4\left[\iota\left(-\frac{\omega}{\alpha}+u
k\right)uV\right.\nonumber\\
&&+u^2V'(1+4\gamma^2V^2)\left.+2V\{a_z+uu'(1+2\gamma^2u^2)\}\right]=0.
\end{eqnarray}

\subsection{Numerical Solutions}

We consider the previous case with the modifications that
$V=\tanh(z),~u=\sqrt{\frac{1-\tanh^2(z)}{1+\tanh^2(z)}}$. Solving
the determinant of the coefficients of the constants $c_1,~c_2$
and $c_3$ in Eqs.(\ref{f4})-(\ref{f6}), we obtain three complex
wave numbers given in Figures 3, 4 and 5.

\begin{figure}
\center \epsfig{file=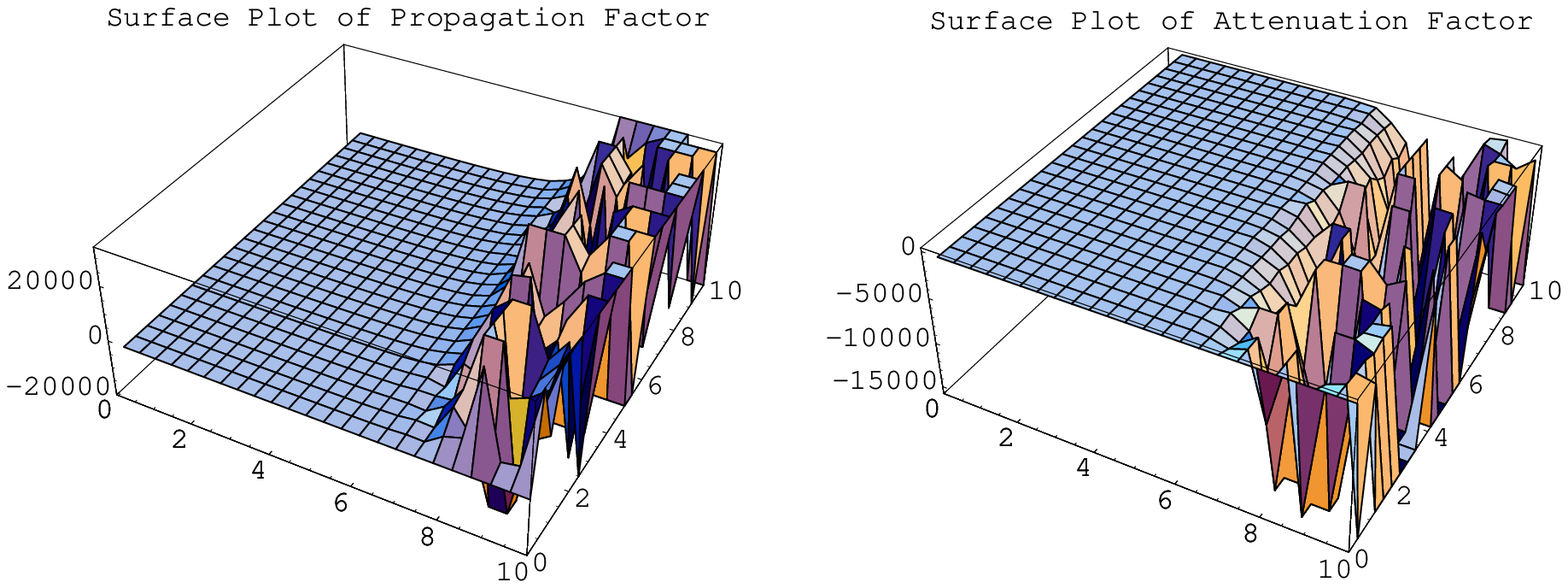,width=1.0\linewidth} \center
\epsfig{file=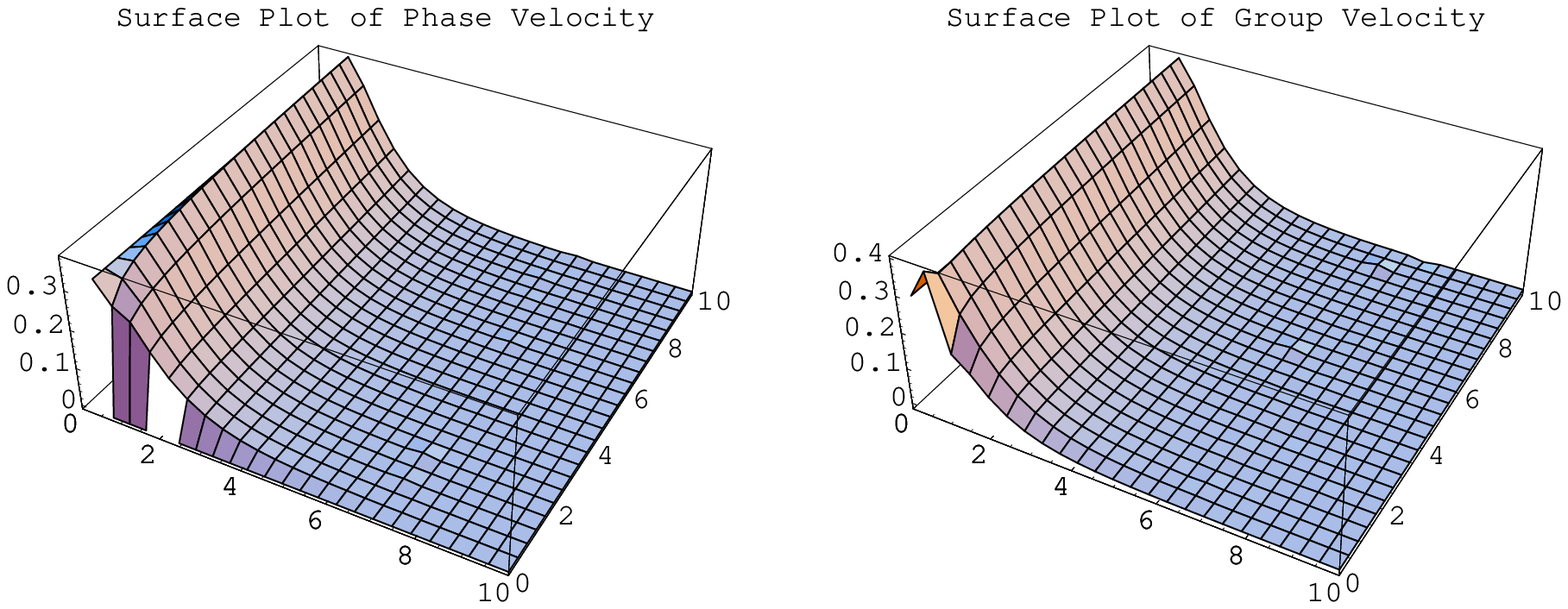,width=1.0\linewidth} \caption{Dispersion is
normal for waves with negligible angular frequencies.}
\end{figure}
In Figure 3, the wave number is infinite at the event horizon. The
propagation factor decreases from the event horizon to $z=0.4$ and
then increases to $z=6.25$. It takes random values in the region
$6.25<z\leq10$. The attenuation factor admits negative values in the
region. The waves damp as they move away from the event horizon
towards $z=0.55$ and grow afterwards up to $z=4.5$. The attenuation
factor assumes random values in the region $4.5<z\leq10$. The phase
velocity of waves is less than the group velocity for the region
$0<z\leq5,~0.1\leq\omega\leq5$ which implies anomalous dispersion.
In the region $0\leq\omega<0.1$, the phase is moving faster than the
group of wavelets and the dispersion is normal.

\begin{figure}
\center \epsfig{file=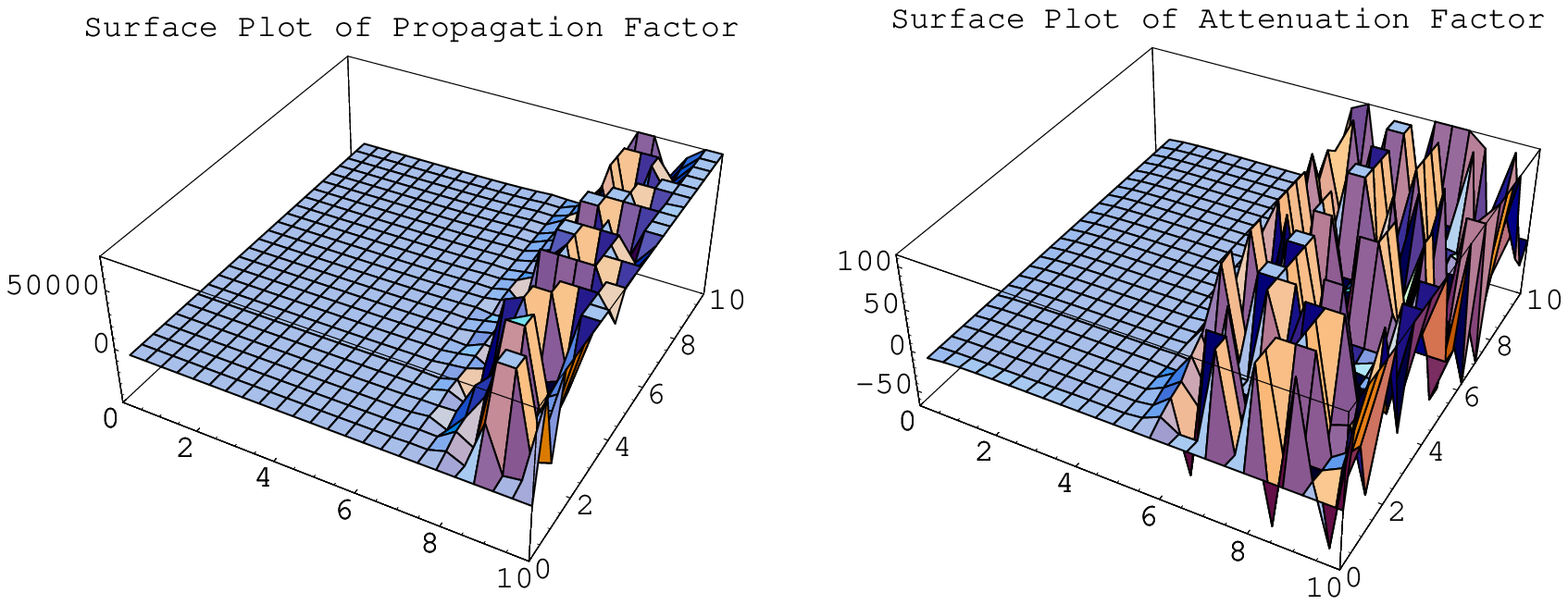,width=1.0\linewidth} \center
\epsfig{file=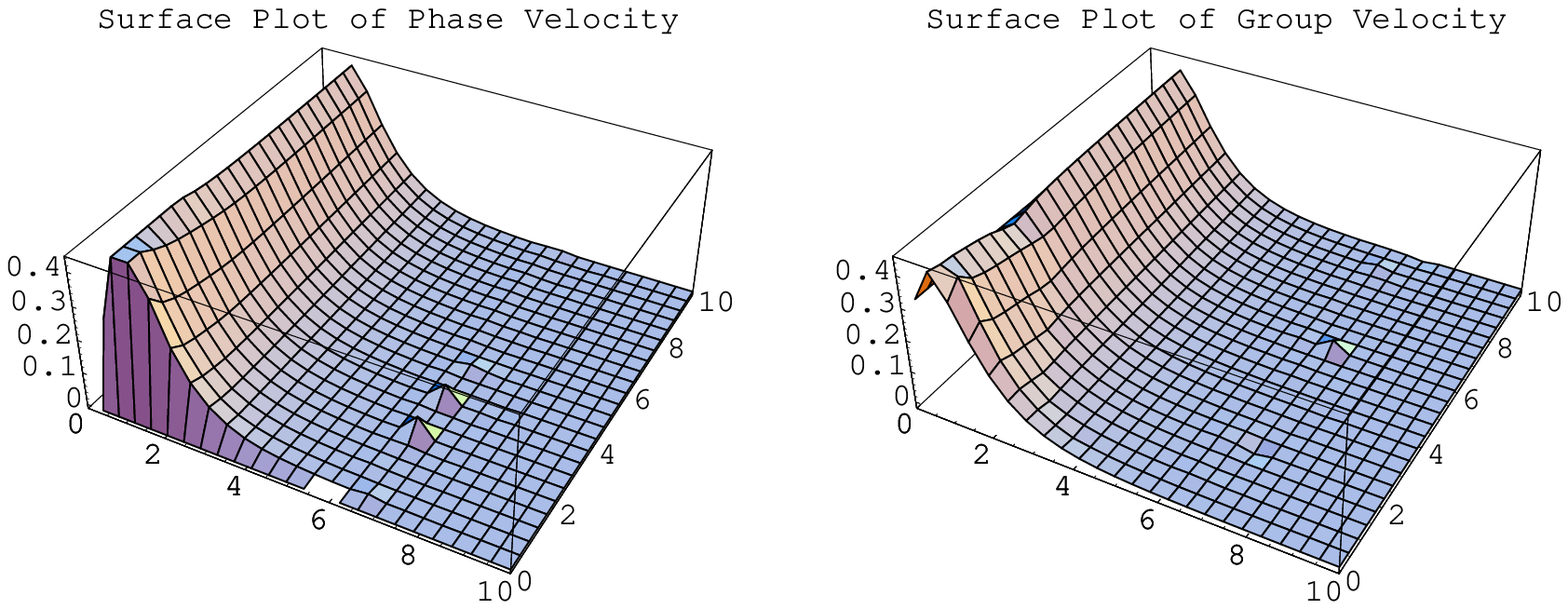,width=1.0\linewidth} \caption{Many small
regions admit normal and anomalous dispersion.}
\end{figure}
Figure 4 shows that no wave lies on the event horizon due to
infinite wave number. The waves propagate rapidly with the increase
in $z$ and $\omega$ in the region $6<z\leq10$. The attenuation
factor increases with the increase in $z$ and $\omega$ in the region
$0<z\leq4.25$ which shows that the waves damp with the increase in
$z$ and $\omega$. The waves with negligible angular frequency do not
show this behavior. The phase velocity exceeds the group velocity
for the regions $0.55\leq z\leq3,~5\leq\omega\leq10$ and $1\leq
z\leq3,~0.15\leq\omega\leq5$ which indicates normal dispersion. The
group velocity is greater than the phase velocity for the region
$0<z<0.55,~5\leq\omega\leq10$ where dispersion is anomalous. In the
regions $0\leq z<1,~0.15\leq\omega\leq5$ and $0<z<0.15$, there lie
random points of normal as well as anomalous dispersion.

\begin{figure}
\center \epsfig{file=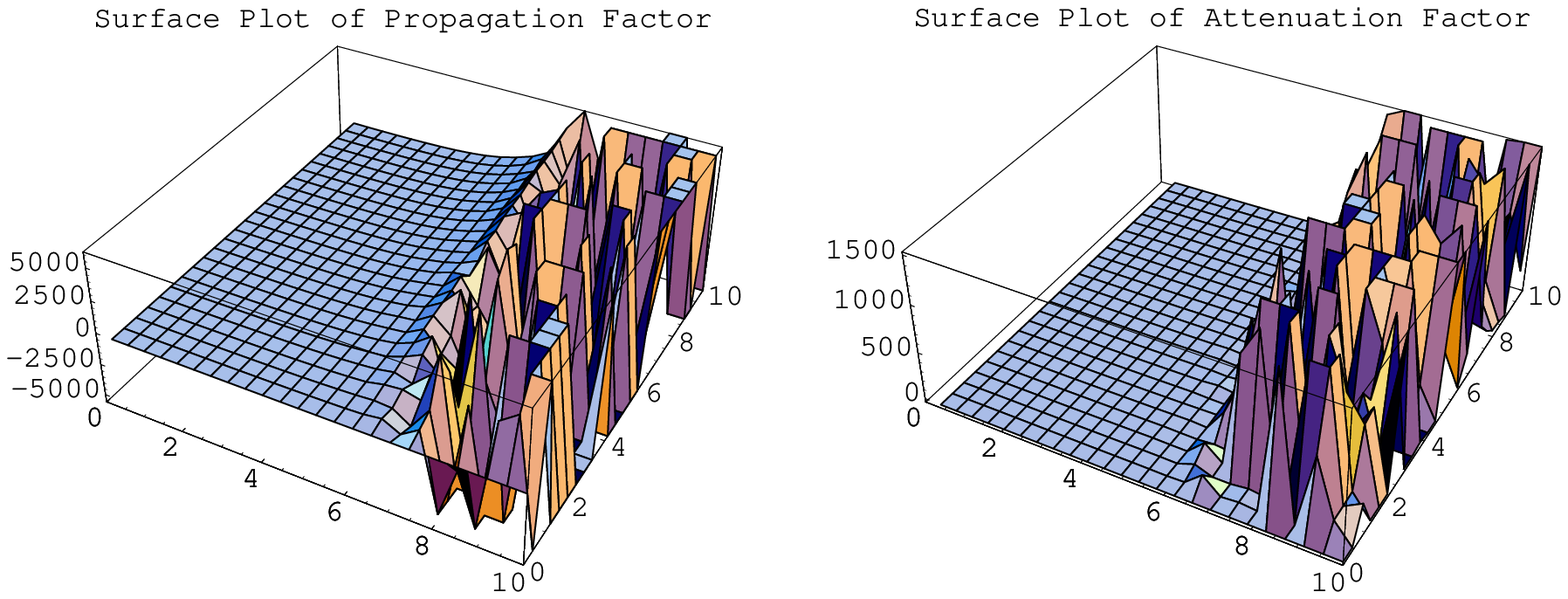,width=1.0\linewidth} \center
\epsfig{file=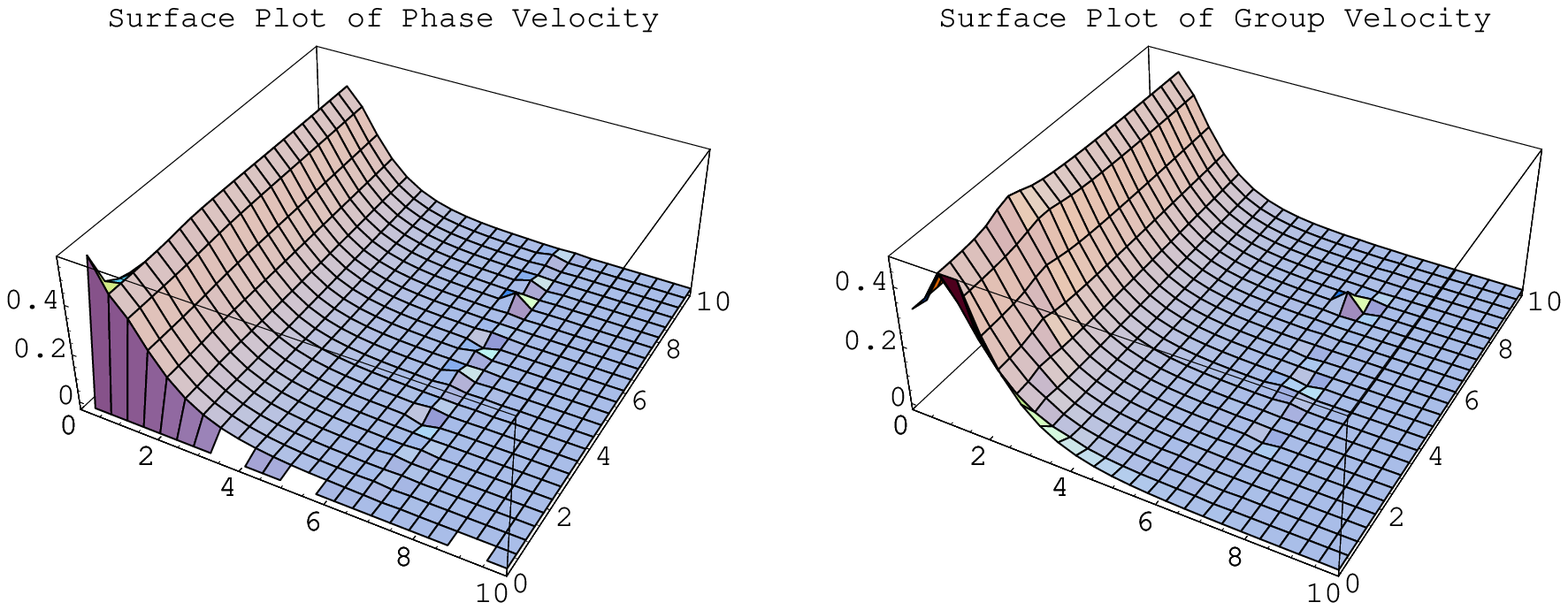,width=1.0\linewidth} \caption{The region
admits normal dispersion for most of the points. The waves with
negligible angular frequencies show anomalous dispersion.}
\end{figure}
In Figure 5, the wave number is infinite at the event horizon. The
propagation factor increases with the increase in $\omega$. It
decreases in $0<z<1$ and increases in $1\leq z\leq6.3$ with the
increase in $z$ and takes random values in the region $6.3<z\leq10$.
In the region $0<z\leq4.6$, the attenuation factor increases with
the increase in $\omega$ and $z$. Thus the waves damp with the
increase in the values of angular frequency and $z$. For
$0<z\leq4.8,~5\leq\omega\leq10$ and $1\leq
z\leq4.8,~1\leq\omega\leq2$, the phase velocity is greater than the
group velocity and the dispersion is normal. Rest of the region
possesses some points of anomalous dispersion but mostly admitting
normal dispersion. The waves with angular frequency between $0$ and
$0.1$ admit anomalous dispersion. The region $4.8<z\leq10$ admits
normal as well as anomalous points of dispersion.

\section{Rotating Magnetized Background}

For harmonic dependence of waves given by Eq.(\ref{b6}), the
Fourier analysed GRMHD Eqs.(5.15)-(5.20) of \cite{U} can be
written as
\begin{eqnarray}{\setcounter{equation}{1}}\label{l1}
&&c_3(\alpha'+\iota k \alpha)-c_2\{(\alpha \lambda)' +\iota
k\alpha\lambda\}+c_5(\alpha V)'\nonumber\\
&&-c_4\{(\alpha u)'-\iota \omega +\iota k \alpha u
\}=0,\\
\label{l2}
&&c_5\left(-\frac{\iota \omega}{\alpha}+\iota k u\right)=0,\\
\label{l3}
&&c_5 \iota k=0,\\
\label{l4} &&c_1\left(-\frac{\iota \omega}{\alpha}+\iota k
u\right)+c_2\left\{-\frac{\iota \omega}{\alpha}\gamma^2 u+\iota
k(1+\gamma^2u^2)+2\gamma^4
u^2VV'\right.\nonumber\\
&&\left.-(1-2\gamma^2u^2)(1+\gamma^2u^2)\frac{u'}{u}\right\}
+c_3\gamma^2\left[\left(-\frac{\iota\omega}{\alpha}+\iota
ku\right)V\right.\nonumber\\
&&\left.+u\{(1+2\gamma^2V^2)V'+2\gamma^2uVu'\}\right]=0,\\
\label{l5}
&&c_1\rho\gamma^2u\{(1+\gamma^2V^2)V'+\gamma^2uVu'\}-\frac{B^2}{4\pi}c_4\{(1-u^2)\iota
k+\frac{\alpha'}{\alpha}(1-u^2)-uu'\}\nonumber\\
&&+c_2\left[-\left(\rho\gamma^4uV-\frac{\lambda
B^2}{4\pi}\right)\frac{\iota\omega}{\alpha}+\iota
ku\left(\rho\gamma^4uV+\frac{\lambda
B^2}{4\pi}\right)+\frac{B^2u}{4\pi\alpha}(\lambda\alpha)'\right.\nonumber\\
&&\left.+\rho\gamma^2\{(1+2\gamma^2u^2)(1+2\gamma^2V^2)
-\gamma^2V^2\}V'+2\rho\gamma^4(1+2\gamma^2u^2)uVu'\right]\nonumber\\
&&+c_3\left[-\left\{\rho\gamma^2(1+\gamma^2 V^2)+
\frac{B^2}{4\pi}\right\}\frac{\iota \omega}{\alpha}+\iota
ku\left\{\rho\gamma^2(1+\gamma^2
V^2)-\frac{B^2}{4\pi}\right\}\right.\nonumber\\
&&\left.+\rho\gamma^4u\{(1+4\gamma^2V^2)uu'+4(1+\gamma^2V^2)VV'\}
-\frac{B^2u\alpha'}{4\pi\alpha}\right]=0,\\
\label{l6}
&&c_1\rho\gamma^2[a_z +u\{(1+\gamma^2 u^2)u'+\gamma^2VuV'\}]\nonumber\\
&&+c_2\left[-\left\{\rho \gamma^2(1+\gamma^2 u^2)+\frac{\lambda^2
B^2}{4\pi}\right\}\frac{\iota \omega}{\alpha}+\left\{\rho
\gamma^2(1+\gamma^2u^2)-\frac{\lambda^2 B^2}{4\pi}\right\}\iota
ku\right.\nonumber\end{eqnarray}\begin{eqnarray}
&&\left.+\left\{\rho\gamma^2\{u'(1+\gamma^2 u^2)(1+4\gamma^2 u^2)
+2u\gamma^2 ((1+2\gamma^2 u^2)VV'+a_z)\}\right.\right.\nonumber\\
&&\left.\left.-\frac{\lambda B^2 u}{4\pi\alpha}(\alpha
\lambda)'\right\}\right]+c_3\left[-\left(\rho \gamma^4uV-
\frac{\lambda B^2}{4\pi}\right)\frac{\iota
\omega}{\alpha}+\left(\rho \gamma^4uV+\frac{\lambda
B^2}{4\pi}\right)\iota ku\right.\nonumber\\
&&\left.+\left\{\rho\gamma^4 \{u^2V'(1+4\gamma^2
V^2)+2V(a_z+uu'(1+2\gamma^2u^2))\}+\frac{\lambda B^2 \alpha'
u}{4\pi\alpha}\right\}\right]\nonumber\\
&&+\frac{B^2}{4 \pi}c_4\left[\lambda (1-u^2)\iota
k+\lambda\frac{\alpha'}{\alpha}(1-u^2)-\lambda uu'+\frac{(\lambda
\alpha)'}{\alpha}\right]=0.
\end{eqnarray}
Equations (\ref{l2}) and (\ref{l3}) imply that $c_5$ is zero which
gives that $b_z$ is zero.

\subsection{Numerical Solutions}

We have assumed the same velocity which we have taken in the
previous section. In addition to this, we consider the same
magnetic field components which we have used in the case of
isothermal fluid in \cite{S2}, i.e.,
$B_x=B\lambda=-\sqrt{\frac{88(1-\tanh^4(z))}{7 \tanh^2(z)}}$ and
$B_z=B=\sqrt{\frac{88}{7}}$.

Under these conditions, when we solve the determinant of
Eqs.(\ref{l1}), (\ref{l4})-(\ref{l6}) by using $c_5=0$, we obtain
a dispersion relation which leads to four complex wave numbers
$k$. These wave numbers are shown in Figures 6-9.

\begin{figure}
\center \epsfig{file=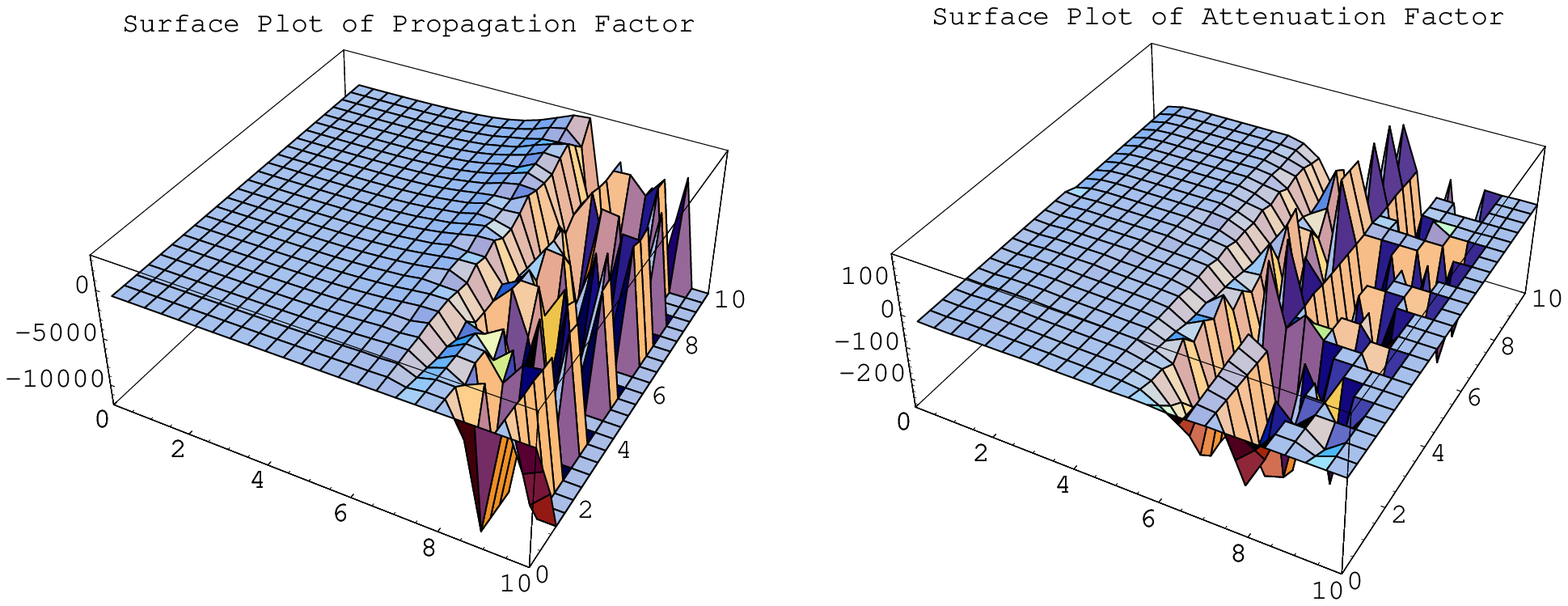,width=1.0\linewidth} \center
\epsfig{file=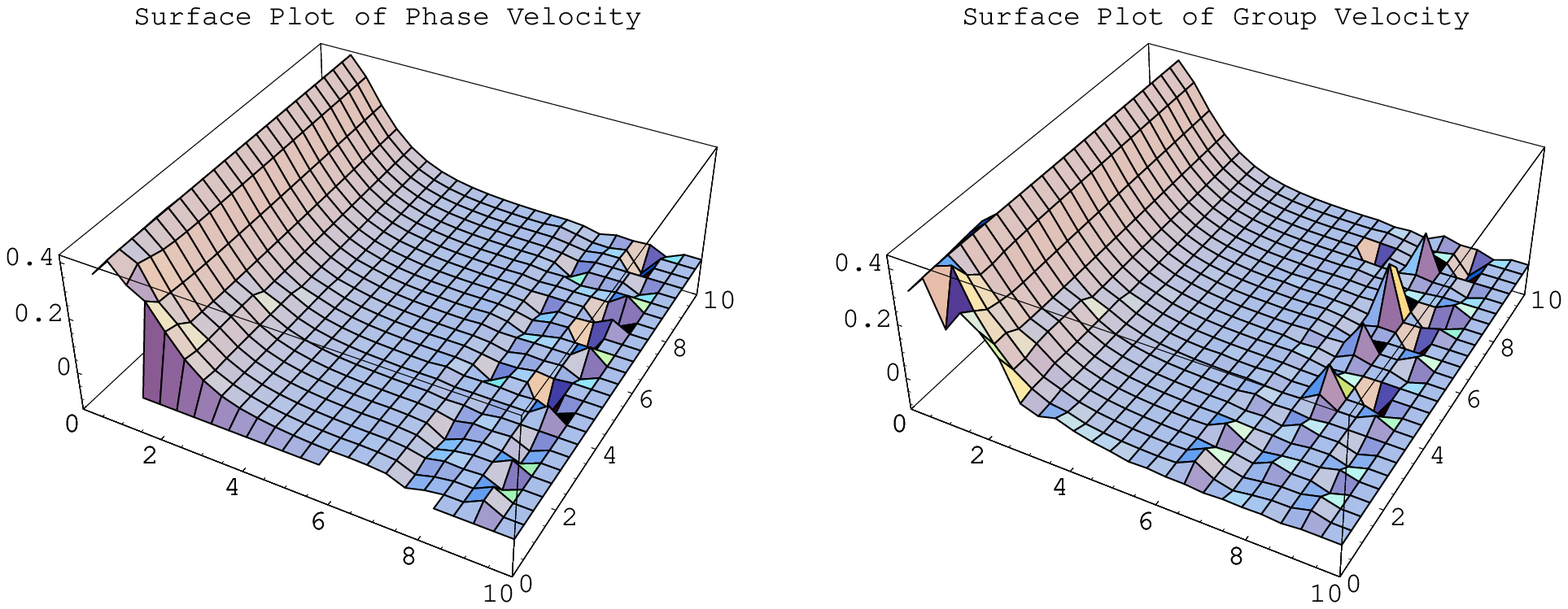,width=1.0\linewidth}\caption{Wave propagation
is less near the event horizon. Anomalous dispersion occur in most
of the region.}
\end{figure}
Figure 6 indicates that the wave number is infinite at the event
horizon. The propagation factor has small values near the event
horizon. It suddenly increases and then decreases as the waves move
towards $z=1$ after which it increases in the region $0<z<6.3$. This
shows that the waves abruptly move faster, then their propagation
decreases and afterwards increases as they move away from the event
horizon in the mentioned region. The propagation of the waves is
fast as we increase their angular frequency which is quite usual.
The propagation factor takes random values in the region $6.3\leq
z\leq10$. The attenuation factor increases when the waves move away
from the event horizon. It decreases in $0.3\leq z\leq0.7$ and then
increases monotonically in the region $0<z\leq4,~1\leq\omega\leq10$.
Thus damping occurs when the waves move away from the event horizon.
The waves grow in the region $0.3\leq z\leq0.7$ after which they
show smooth damping.

The phase velocity is less than the group velocity for the waves
near the event horizon which shows normal dispersion. As the waves
move away from the event horizon, the group velocity overcomes the
phase velocity and anomalous dispersion holds. The waves with low
angular frequency admit anomalous dispersion. As the angular
frequency increases, the waves show normal dispersion near the
event horizon but as these waves move away, they disperse
anomalously. Thus these waves cannot transfer any information out
of the region.

\begin{figure}
\center \epsfig{file=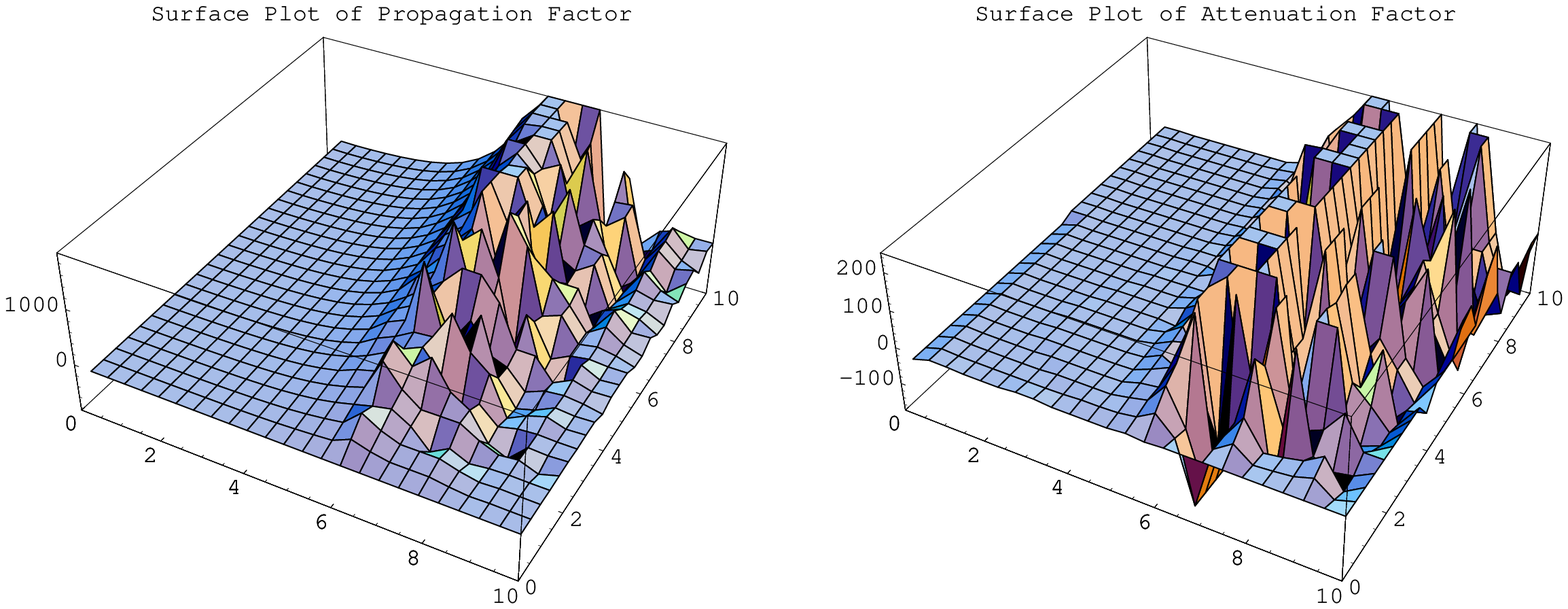,width=1.0\linewidth} \center
\epsfig{file=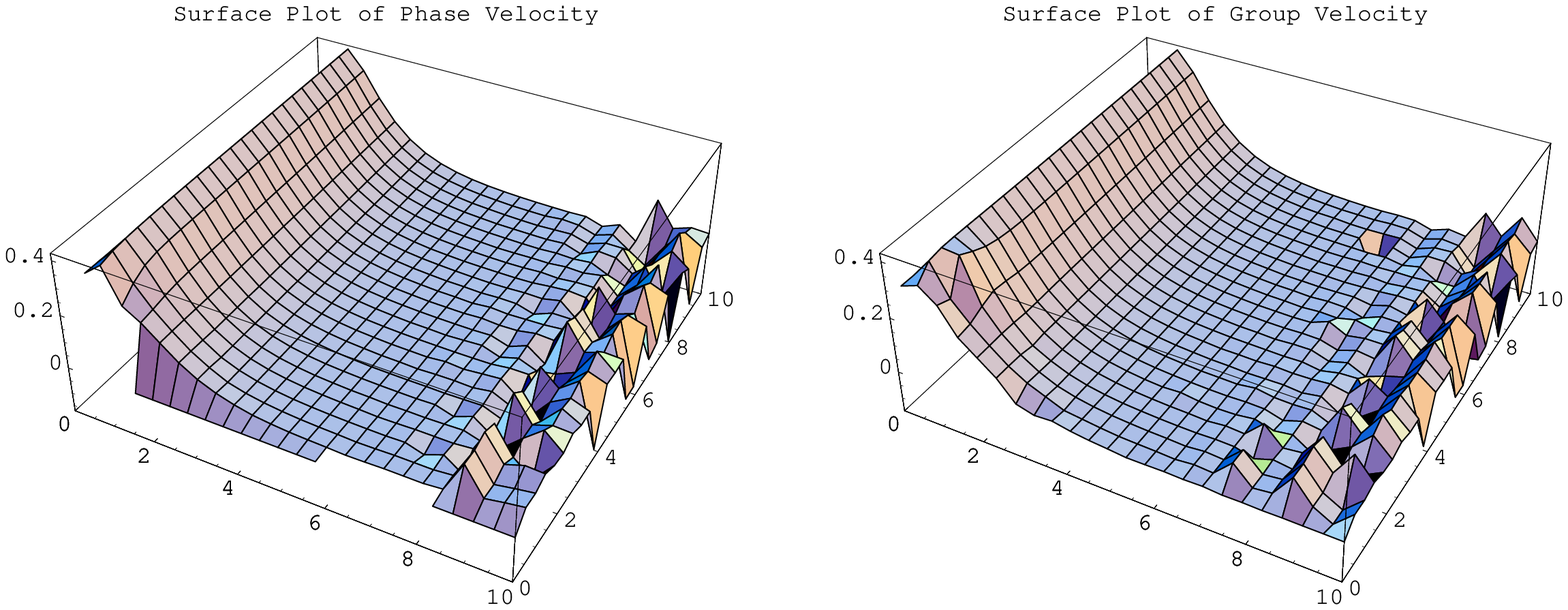,width=1.0\linewidth} \caption{Dispersion is
random in most of the region. Negative wave propagation region is
observable.}
\end{figure}

Figure 7 demonstrates that the wave number is infinite at $z=0$
and thus no wave exists there. The propagation factor decreases
abruptly when the waves move away from the event horizon. This
factor increases with an increase in the value of $z$ and
$\omega$. This means that the wave propagation abruptly decreases
when the waves move away from the neighborhood of the event
horizon of the Schwarzschild black hole. Then their propagation
increases gradually. The waves with higher angular frequency
propagate more rapidly in the environment than the waves with
lower angular frequency. The attenuation factor decreases when
moving away from the event horizon till $z=0.5$. It then increases
a little and decreases gradually in the region
$0<z\leq2,~2\leq\omega\leq10$. The waves grow as they move away
from the event horizon, a little damping occurs after which they
grow again in this region. In the rest of the region, the growth
and damping occur randomly.

The group velocity is greater than the phase velocity in the region
$2\leq z\leq10,~4\leq\omega\leq10$ which indicates anomalous
dispersion. The rest of the region shows random points of
dispersion. In the region $8\leq z\leq10$, the phase as well as the
group velocity admit the negative values.

\begin{figure}
\center \epsfig{file=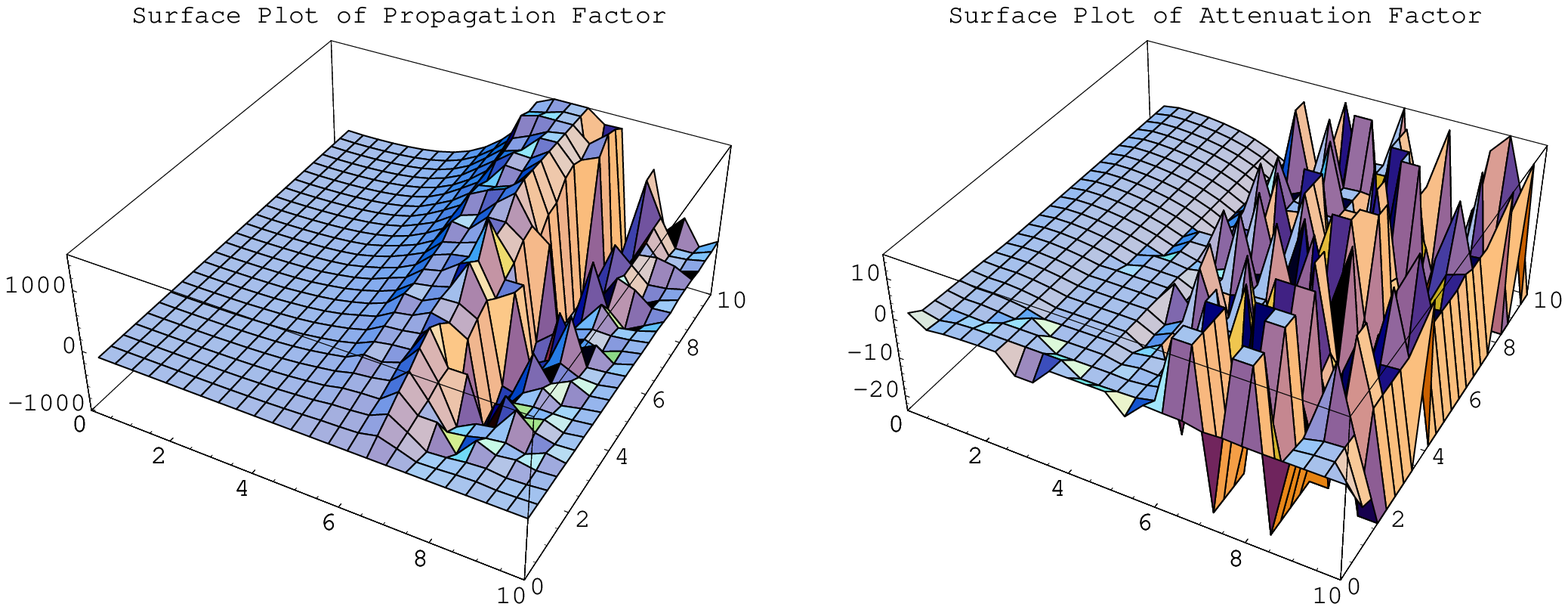,width=1.0\linewidth} \center
\epsfig{file=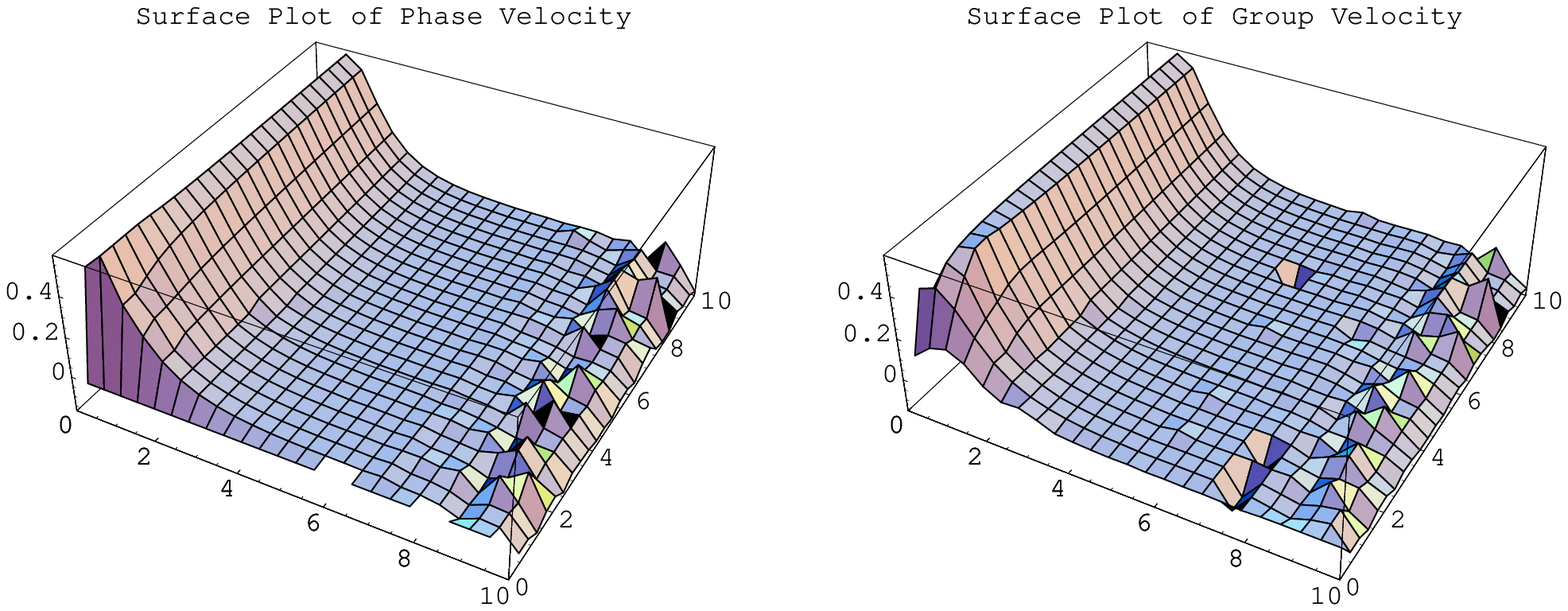,width=1.0\linewidth} \caption{Small regions
of normal as well as anomalous dispersion are found. Negative wave
propagation region occur.}
\end{figure}

Figure 8 indicates that when the waves move away from the event
horizon, the propagation factor decreases towards $z=1$ and
increases gradually towards $z=6$. In the region $6<z\leq10$, the
propagation factor attains random values and in the region
$0<z\leq2,~2\leq\omega\leq10$, the attenuation factor decreases as
$z$ increases. This indicates that the waves damp with the increase
in $z$ in this small region. In the rest of the region, the
attenuation factor admits random values and thus the growth and
damping of waves occur at random.

The phase velocity is greater than the group velocity for most of
the points in the regions $2\leq z\leq3,~0.5\leq\omega\leq4$ and
$3\leq z\leq4,~1\leq\omega\leq6$ which indicates normal dispersion.
The group velocity exceeds the phase velocity and the dispersion is
normal in the regions (i) $1\leq z\leq2,~2\leq\omega\leq10$, (ii)
$2\leq z\leq3,~4\leq\omega\leq10$ and (iii) $3\leq
z\leq3.5,~6\leq\omega\leq10$. The rest of the region contains points
of normal as well as anomalous dispersion. The region $6\leq
z\leq10$ takes some points at which the phase as well as the group
velocity have negative values.

\begin{figure}
\center \epsfig{file=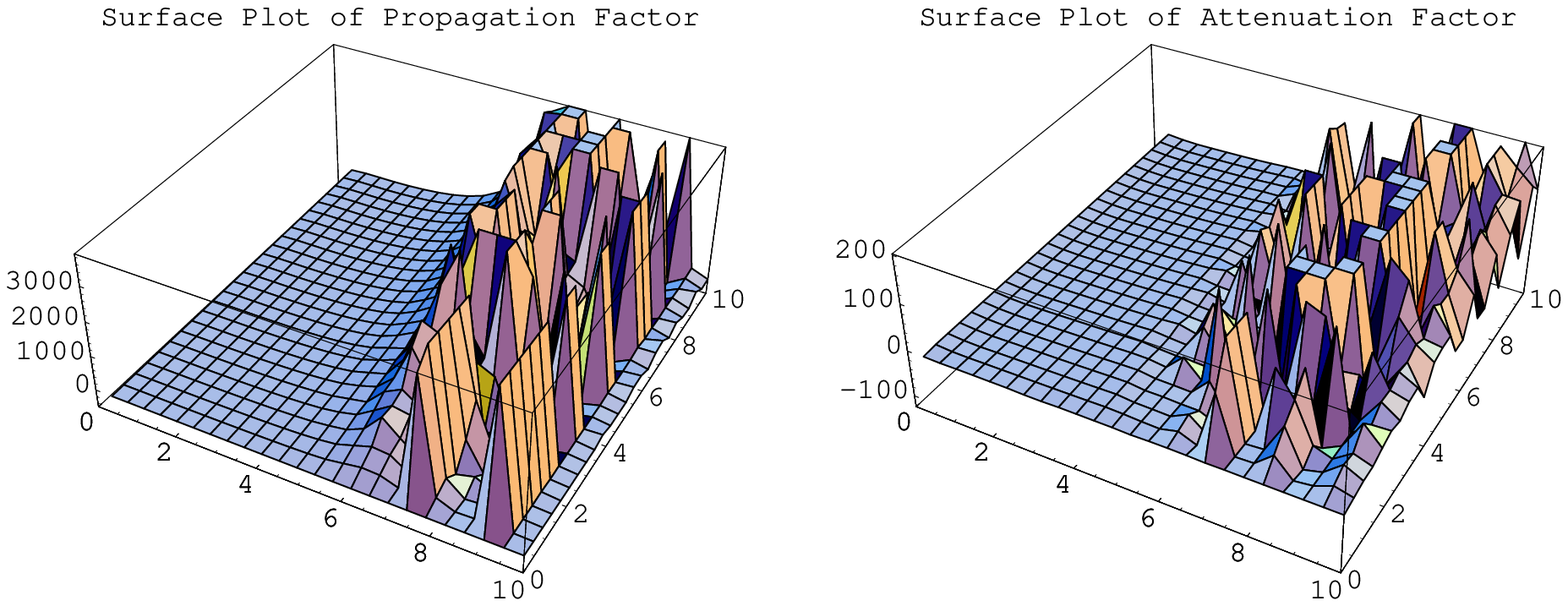,width=1.0\linewidth} \center
\epsfig{file=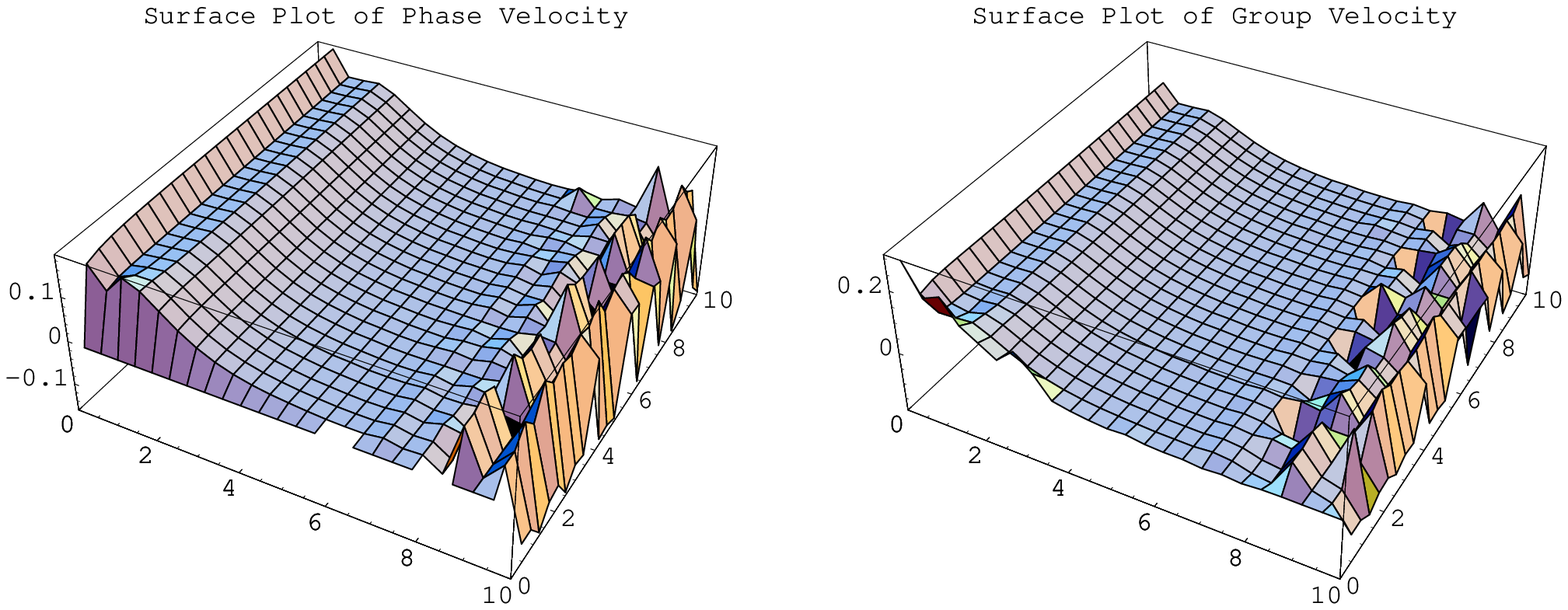,width=1.0\linewidth} \caption{Most of the
regions contains random points of dispersion. Region with negative
phase velocity propagation is also present.}
\end{figure}

In Figure 9, the wave number is infinite at $z=0$. The propagation
factor decreases at once on going away from the event horizon. It
increases gradually when the value of $z$ as well as the angular
frequency increase in the region. This means that the propagation
of waves decreases abruptly when they move away from the event
horizon. and increases when the waves move faster. The waves with
higher angular frequency propagate faster than the waves with
lower angular frequency. The propagation factor assumes random
values in the region $6\leq z\leq10$. The attenuation factor has
less variations in the region $0<z\leq4$ and shows increasing
behavior as the value of $z$ and $\omega$ increases. This means
that the waves damp either when they move away from the event
horizon or when their angular frequency is increased. The
variation of this factor is much more in the region $4<z\leq10$.

In the regions (i) $0<z<3,~4\leq\omega\leq10$, (ii) $1\leq
z<4,~2\leq\omega\leq4$ and (iii) $0<z\leq2.5,~2\leq\omega\leq4$, the
phase velocity is greater than the group velocity and the dispersion
is normal. The regions $3\leq z\leq4.4,~4\leq\omega\leq10$ and
$2.5\leq z\leq5,~2\leq\omega\leq4$ possess anomalous dispersion due
to the fact that the group velocity is greater than the phase
velocity. The regions $0\leq\omega\leq2$ and $4.4<z\leq 10$ admit
some points with normal and some with anomalous dispersion. In the
region $5<z\leq 10$, the phase velocity takes negative values at
some points and hence the region contains points with negative phase
velocity propagation.

\section{Outlook}

In this paper, we work out the complex wave numbers near the event
horizon of the Schwarzschild black hole. For this purpose, we have
taken the planar analogue of the Schwarzschild spacetime with
non-rotating and rotating, non-magnetized and magnetized
backgrounds. We have investigated \cite{U} real dispersion relations
for the same backgrounds by using Rindler coordinates. The Rindler
coordinates give best approximation of the Schwarzschild spacetime
near the event horizon whereas this planar analogue is valid for the
whole accretion cloud around the Schwarzschild event horizon. The
properties of the surrounding plasma are based on the complex wave
numbers.

The summary of the results obtained is outlined below:
\begin{itemize}
\item The wave number is
infinite at the horizon which supports the well-known fact that no
information can be extracted from a black hole.
\item There is no difference in the dispersion relations whether
the plasma is magnetized or not, in the case of non-rotating
background because the gravity produces no perturbations in the
magnetic field. The fluid is dispersed normally in only one case
(Figure 2) which allows the waves to pass through this region. The
waves show anomalous dispersion in the other case.
\item The rotating background exhibits negative phase
velocity propagation regions far away from the black hole event
horizon. It seems that the black hole's gravity ceases the waves
to admit the negative phase velocity propagation region. For the
region with larger values of $z$, the dispersion is found to be
normal at some points and anomalous at some other points. This
creates a doubt whether all the waves can get out of this region.
\item There is only one case (Figure 5) in which normal dispersion
occurs at most of the points in rotating non-magnetized background.
\item For the rotating magnetized background, many small regions
which admit normal dispersion are found. These regions are
surrounded by the regions of anomalous dispersion and hence no
information can be extracted from these regions.
\end{itemize}

The comparison of the results with those obtained from the
previous literature is given as follows:
\begin{itemize}
\item When we compare our results to that of
\cite{U}, we find that in non-rotating background we have detected
the anomaly of waves which was not recognized previously. In the
case of rotating magnetized background we obtain the negative
phase velocity propagation region far away from the event horizon
whereas in \cite{U}, the whole region admits this property.
\item In the case of non-rotating isothermal plasma \cite{S1}, the
dispersion is anomalous in all the cases whereas in cold plasma
the region with normal dispersion is also found. It can be deduced
that the plasma pressure ceases the chances of waves to come out
of the region in the neighborhood of the event horizon in pure
Schwarzschild background.
\item In the case of rotating magnetized isothermal plasma \cite{S2},
the medium admits normal dispersion of waves in plasma surrounding
the event horizon for the waves admitting high angular frequency.
For the same case of cold plasma, the regions with normal dispersion
are enclosed by the regions admitting anomalous dispersion of the
waves. For the restricted Kerr black hole, it can be observed that
the pressure reduction ceases the chances of waves to pass through
the region.
\end{itemize}

We conclude that no information can be extracted from the
Schwarzschild event horizon. There are chances to collect the
information from the exterior of the event horizon of the
Schwarzschild black hole for the case of cold plasma. In the case of
the restricted Kerr geometry, the cold plasma does not allow the
waves to pass through in most of the cases. However, there are
little chances for the waves to move out of the accretion disk.

\vspace{0.4cm}

{\bf Acknowledgment}

\vspace{0.4cm}

We appreciate the Higher Education Commission Islamabad, Pakistan,
for its financial support during this work through the {\it
Indigenous PhD 5000 Fellowship Program Batch-II}.


\vspace{0.5cm}

\begin{thebibliography}{99}
\bibitem{P}Petterson, J.A.: Phys. Rev. \textbf{D10}(1974)3166.

\bibitem{ADM}Arnowitt, R., Deser, S. and Misner, C.W.: \textit{Gravitation: An
Introduction to Current Research} ed. Witten, L. (Wiley, New York,
1962).

\bibitem{SJ} Stachel, J.: Acta. Phys. Polon.
\textbf{35}(1969)689.

\bibitem{SY}Smarr, L. and York, J.W., Jr.: Phys. Rev.
\textbf{D17}(1978)2529.

\bibitem{Y}York, J.W., Jr.: \textit{Sources of Gravitational Radiation} ed.
Smarr, L. (Cambridge University Press, Cambridge, 1979).

\bibitem{STW}Smarr, L., Taubes, C. and Wilson, J.R.: \textit{Eassays in General
Relativity, A Festschrift for Ibraham Taub}, ed. Tipler, F.
(Academic Press, New York, 1980).

\bibitem{ESW}Evans, C.R., Smarr, L.L. and Wilson, J.R.: \textit{Astrophysical
Radiation Hydrodynamics} ed. Norman, M. and Winkler, K.H. (Reidel,
Dordrecht, 1986).

\bibitem{TM1}Thorne, K.S. and Macdonald, D.A.: Mon. Not. R. Astron. Soc.
\textbf{198}(1982)339.

\bibitem{TM2}Thorne, K.S. and Macdonald, D.A.: Mon. Not. R. Astron. Soc.
\textbf{198}(1982)345.

\bibitem{TPM} \textit{Black Holes: The Membrane Paradigm}
eds. Thorne, K.S., Price, R.H. and Macdonald, D.A. (Yale
University Press, New Haven, 1986).

\bibitem{HT}Holcomb, K.A. and Tajima, T.: Phys. Rev. \textbf{D40}(1989)3809.

\bibitem{H}Holcomb, K.A.: Astrophys. J. \textbf{362}(1990)381.

\bibitem{De}Dettmann, C.P., Frankel, N.E. and Kowalenko, V.:  Phys. Rev.
\textbf{D48}(1993)5655.

\bibitem{SK}Sakai, J. and Kawata, T.: J. Phys. Soc. Jpn. \textbf{49}(1980)747.

\bibitem{Kh}Khanna, R.: Mon. Not. R. Astron. Soc. \textbf{294}(1998)673.

\bibitem{Z1}Zhang, X.-H.: Phys. Rev. \textbf{D39}(1989)2933.

\bibitem{Z2}Zhang, X.-H.: Phys. Rev. \textbf{D40}(1989)3858.

\bibitem{BH1}Buzzi, V., Hines, K.C. and Treumann, R.A.: Phys. Rev.
\textbf{D51}(1995)6663.

\bibitem{BH2}Buzzi, V., Hines, K.C. and Treumann, R.A.: Phys. Rev.
\textbf{D51}(1995)6677.

\bibitem{U} Sharif, M. and Sheikh, U: Gen. Relat. Gravit. \textbf{39}(2007)1437.

\bibitem{S1} Sharif, M. and Sheikh, U: \textit{Effects of Schwarzschild Geometry
on Isothermal Plasma Wave Dispersion}, submitted for publication.

\bibitem{S2} Sharif, M. and Sheikh, U: \textit{Effects of Rotating Background on
Isothermal Plasma Wave Dispersion in Schwarzschild Geometry},
submitted for publication.

\bibitem{Ac} Achenbach, J.D.: \textit{Wave Propogation in Elastic Solids}
(North-Holland Publishing Company, Oxford, 1973).

\end{thebibliography}
\end{document}